\documentclass[lettersize,journal]{IEEEtran}
\usepackage{amsmath,amsfonts}
\usepackage{algorithmic}
\usepackage{algorithm}
\usepackage{array}
\usepackage[caption=false,font=normalsize,labelfont=sf,textfont=sf]{subfig}
\usepackage{textcomp}
\usepackage{stfloats}
\usepackage{url}
\usepackage{verbatim}
\usepackage{graphicx}
\usepackage{cite}
\usepackage{pifont}
\newcommand{\cmark}{\ding{51}}%
\newcommand{\xmark}{\ding{55}}%

\usepackage{caption}
\usepackage{multicol}
\usepackage{multirow}

\begin{document}


\title{Interactive singing melody extraction based on active adaptation\\
{\thanks{This work was supported by grant no. PB/EE/2021128B from Prasar Bharti.}}}

\author{Kavya Ranjan Saxena, Vipul Arora,~\IEEEmembership{Member,~IEEE}\\{kavyars@iitk.ac.in, vipular@iitk.ac.in}\\{Department of Electrical Engineering}\\{Indian Institute of Technology Kanpur, India}}

\markboth{Journal of \LaTeX\ Class Files,~Vol.~14, No.~8, August~2021}%
{Shell \MakeLowercase{\textit{et al.}}: A Sample Article Using IEEEtran.cls for IEEE Journals}

\maketitle
\begin{abstract}
Extraction of predominant pitch from polyphonic audio is one of the fundamental tasks in the field of music information retrieval and computational musicology. To accomplish this task using machine learning, a large amount of labeled audio data is required to train the model. However, a classical model pre-trained on data from one domain (source), e.g., songs of a particular singer or genre, may not perform comparatively well in extracting melody from other domains (target). The performance of such models can be boosted by adapting the model using very little annotated data from the target domain. In this work, we propose an efficient interactive melody adaptation method. Our method selects the regions in the target audio that require human annotation using a confidence criterion based on normalized true class probability. The annotations are used by the model to adapt itself to the target domain using meta-learning. Our method also provides a novel meta-learning approach that handles class imbalance, i.e., a few representative samples from a few classes are available for adaptation in the target domain. Experimental results show that the proposed method outperforms other adaptive melody extraction baselines. The proposed method is model-agnostic and hence can be applied to other non-adaptive melody extraction models to boost their performance. Also, we released a Hindustani Alankaar and Raga (HAR) dataset containing 523 audio files of about 6.86 hours of duration intended for singing melody extraction tasks.
\end{abstract}


\section{Introduction}
Extracting sining melody from polyphonic audio is a fundamental and important task in the music information retrieval field. The aim is to extract the pitch of the dominant singing voice from polyphonic audio. There are many downstream applications of melody extraction, including music recommendation \cite{musicrecom}, cover song identification \cite{coversong}, music generation \cite{musicgen}, and voice separation \cite{voicesep}.

Machine learning methods generally use supervised learning that involves training a model on source domain data and testing/deploying on target domain data. These models provide excellent performance when sufficiently large annotated data is available in the source domain and the data distribution of the target domain is approximately same as that of source domain. But the performance degrades when these models are applied to different target domains which may vary in data distribution as compared to that of source domain~\cite{patch-basedcnn}. This is called domain shift. For example, a model trained on songs of a particular singer or genre (source domain), may not perform comparatively well in extracting melody for a different singer or genre (target domain). 
In this paper, we study the effect of domain shift on melody extraction and propose methods to tackle this problem. The performance degradation by domain shift can be avoided by adapting on a very little annotated data in the target domain. This is referred to as domain adaptation\cite{da1}. In this paper, we propose a novel model-agnostic active-meta-learning-based domain adaptation technique that is a combination of active-learning~\cite{alsurvey} and meta-learning~\cite{maml-finn}. Given a spectrogram of an audio in the target domain, the model uses active-learning to select those frames of the spectrogram where it is least confident and marks those frames for the human annotator to annotate. Once the human annotator provides the melody annotations for those frames, the model uses meta-learning to adapt its parameters. In this way, the model adapts to the target domain. One major application of the proposed domain adaptation technique is to obtain precise melody annotations for a large corpus of unlabelled audio with minimum human effort.

In this paper, melody extraction is treated as a classification problem, where the pitch values are binned into a fixed number of pitch classes. This leads to high class imbalance in the data and the trained model would be severely biased to the majority classes. Generally meta-learning algorithm~\cite{maml-finn} is used for domain adaptation in few-shot learning~\cite{fsl1} setting where each class is represented with small number of examples, but in case of melody extraction we do not have representative samples for each class. Hence we modify the vanilla meta-learning approach to handle the severe class imbalance seen in the task at hand. In this paper we follow optimization-based meta-learning algorithm which is robust in its ability to quickly adapt to a few samples from a new target domain. 

The main contributions of this work are:
\begin{itemize}
    \item A comprehensive study on the problem of domain shift in polyphonic melody extraction. 
    \item A novel meta-learning-based adaptation approach to handle severe class imbalance in classification.
    \item A novel interactive domain adaptation method that combines active-learning with meta-learning.
    \item We apply the above methods to effectively tackle the problem of domain shift in melody extraction. To the best of our knowledge, no work on domain adaptation for singing melody extraction has been done in the past.
    \item We release a new dataset named Hindustani Alankaar and Raga (HAR) dataset for singing melody extraction task. The dataset will be accessible using the link \url{https://zenodo.org/record/8252222} 
\end{itemize}

\section{Related Works}
\label{relatedworks}
\subsection{Existing works on melody extraction}
An earlier attempt at extracting melody from polyphonic audio is inclined toward signal processing methods\cite{sp1}. Due to the presence of accompaniments in polyphonic audio, often the first harmonic gets distorted. So, instead of conventional methods, which tracked pitch values, a robust harmonic comb tracking approach\cite{aroraonline} is proposed that focuses on the strong higher harmonics. This method extracts the melody in real time, which has applications in query-based music search. In \cite{salamonphd}, many other non-deep learning approaches such as salience-based and source separation based approaches are summarized along with various applications and challenges in the field of polyphonic melody extraction. With the advances in the field of deep learning, various neural network-based methods have been proposed to extract melody from polyphonic audios. One such work by Lu et al. \cite{aud-sym-tl} uses a deep convolutional neural network (DCNN) with dilated convolution as the semantic segmentation tool. The candidate pitch contours on the time-frequency image are enhanced by combining the spectrogram and cepstral-based features. Another work by Bittner et al.\cite{deepsalience} describes a fully convolutional neural network for learning salience representations for estimating fundamental frequencies. Another proposed encoder-decoder architecture by Hsieh et al.\cite{en-decoder} is used to estimate the presence of melody line and improve the performance by independently recognizing the voiced and unvoiced frames.  To improve the performance of these networks, varied musical and structural context is required. For example, classification tasks\cite{jdc} are used to jointly detect the voiced and unvoiced frames. Attention networks\cite{attention} are used to further capture the relationship between frequencies.

All the above deep-learning based methods employ a standard supervised learning approach for melody extraction from polyphonic audio involving training a model on source domain data and testing on target domain data. The models are not adapted to annotated data in the target domain. In this paper, we propose a novel domain adaptation algorithm that is model-agnostic and can be applied to the non-adaptive models to improve their performance.  

\subsection{Existing domain adaptation techniques}
Domain adaptation techniques are used to minimize the domain shift between the source and target domain by adapting on a few annotated data from the target domain. Such type of adaptation is called supervised domain adaptation (SDA). The other types of domain adaptation techniques are unsupervised~\cite{uda1} and semi-supervised\cite{ssda1} domain adaptation. In this paper, we focus only on SDA. 
Tzeng et al.~\cite{sdt} introduced an auxiliary adversarial task of domain classification to learn domain invariant embeddings. Additionally, they match the softmax output for a sample of a particular class from target domain with the mean softmax output for all samples of that class in the source domain. Motiian et al. ~\cite{unifiedsda} suggested to use contrastive loss to minimize the distance between the samples of same class from source and target domains and simultaneously penalize the distance of samples from different classes from source and target domain. This idea is further extended in neural embedding matching~\cite{neuralembed} that adds a constraint, encoded using using graph-embedding techniques to preserve the local geometry of data across domains. Xu et al.~\cite{dsne} proposed to use stochastic neighbourhood embedding that uses modified-Hausdorff distance for supervised domain adaptation.


\subsection{Existing works on Meta-learning}
The most popular definition of meta-learning is learning to learn, that focuses on improving the learning algorithm over multiple learning episodes. In meta-learning, an inner learning algorithm solves a task such as image classification~\cite{imagenet} that is defined by a dataset and objective. Further, the outer learning algorithm updates inner learning algorithm to improve the outer objective that can be either generalization or learning speed of the inner learning algorithm. The existing works on meta-learning are divided into three categories, namely, metric-based~\cite{fsl1}, model-based~\cite{fsl2} and optimization-based~\cite{modelini} learning.
Optimization-based methods include approaches where the inner-level task is explicitly formulated as an optimization problem. These methods primarily aim to obtain meta-knowledge that may be utilized to enhance the optimization performance. A well-known example is MAML~\cite{maml-finn}, which aims to learn good initialization parameters in such a way that a few iterations of inner learning algorithm yields a classifier that performs well on validation data.

All the above methods are used in few-shot learning setting. In this paper, we modify MAML\cite{maml-finn} such that it can handle sparse classes or class imbalance. For validation of the approach we apply it to melody extraction problem. MAML focuses on learning good initialization parameters for a model trained on the source domain so that it quickly adapts to the target domain with little training data.

\subsection{Existing works on active learning}
Active learning is the technique by which the model aims to select the most useful samples from a pool of unlabeled samples and provide them to the annotator for labeling. It is done to reduce the cost of labeling by simultaneously maintaining high performance of the model. There are three major approaches of selection of samples from unlabeled samples namely, uncertainty-based approach, diversity-based approach and expected model change. The uncertainty-based approach~\cite{uncen1}\cite{uncen2} determines the amount of uncertainty and measures it to choose uncertain samples. The diversity-based approach~\cite{diversity1}\cite{diversity2} chooses a variety of samples that reflect the entire distribution of the unlabeled samples. The expected model change~\cite{expected1}\cite{expected2} refers to the selection of data points that would result in the most significant change to the current model parameters or outputs, assuming knowledge of their corresponding labels. The uncertainty of a sample is determined by the probability of a predicted class~\cite{unal1} or the entropy of class posterior probabilities~\cite{ecp}. Lewis et al.~\cite{unal1} used only one classifier to select those samples where the classifier is least confident. Gal et al.~\cite{gal} employ the Monte Carlo Dropout technique~\cite{mcmc} to derive uncertainty estimates from deep networks by conducting several forward passes.

In this work, we use normalized true class probability of the sample as the uncertainty measure for selecting samples in active learning approach.


\begin{figure}
 \centering
 \includegraphics[width=7.5cm, height=5cm]{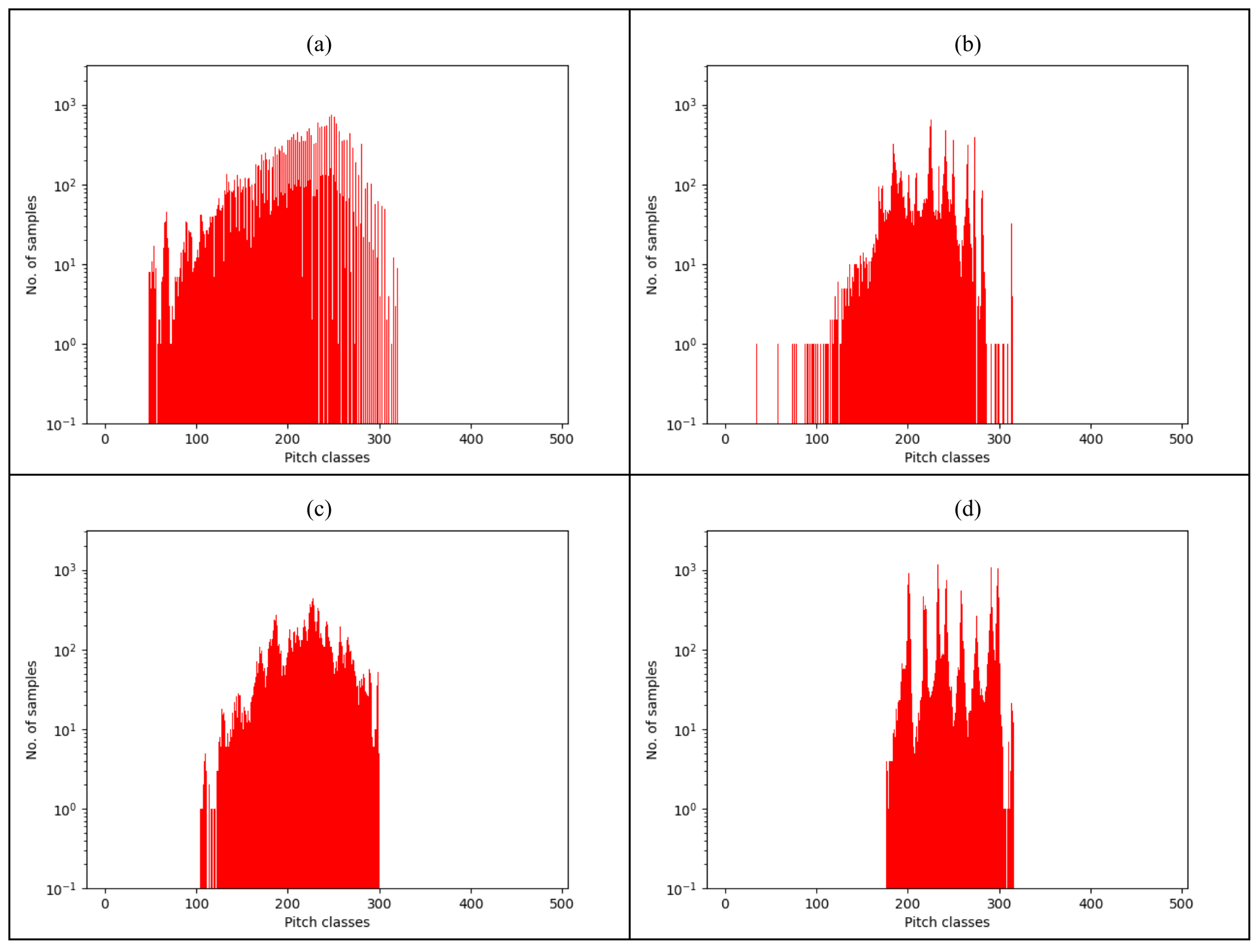}
 \caption{Class imbalance in the (a) source domain (MIR1K) and different target domains (b) ADC2004, (c) MIREX05 and (d) HAR. Class 0 represents the non-voiced class and classes 1-506 represent voiced pitch classes ranging from A1(55 Hz) to B6(1975.7 Hz). Samples corresponding to non-voiced class are not shown as they are highly disproportionate in comparison to the voiced classes.} 
 \label{fig_class_im}
\end{figure}

\section{Methodology} \label{method}
The audio waveforms are merged into a mono channel and then downsampled to 8kHz. Since the audios are of different duration, we have divided the audios into 5-second chunks. By using short-time Fourier transform, we calculate the magnitude spectrogram of the audio chunks. The spectrogram of dimension $F\times M$ is calculated using a 1024-point Hanning window and a hop size of 10ms, where $F$ is the number of frequency bins and $M$ is the number of time frames. The spectrogram is given as an input to the model such that each time frame $M$ is classified into one of the $C=506$ pitch classes, including a non-voiced class. The voiced pitch classes range from A1 (55 Hz) to B6 (1975.7 Hz) with a resolution of 1/8 semitone.

\subsection{Pre-training} \label{sub:pt}
Let the source training dataset be $D_1^S=\{(X_i,Y_i)\}_{i=1}^{I}$, where $X_i$ is the spectrogram of dimension $F\times M$ and $Y_i\in\{0,1\}^{C\times M}$ is a one-hot vector over $c$ classes for every time frame $M$. The time frames of the spectrogram corresponding to the non-voiced class could dominate the voiced classes, thus introducing class imbalance detailed in Fig.~\ref{fig_class_im}. 

We pre-train the base model $f_{[\phi,\theta]}$ on the $D_1^S$ dataset. Here, $\phi$ and $\theta$ are the feature extractor layers and classifier layer, respectively. Initially, the trainable parameters $\phi$ and $\theta$ are randomly initialized. With the spectrogram $X_i$ as the input, the base model $f_{[\phi,\theta]}$ predicts an output distribution $\hat{Y}_i$
of dimension $506 \times M$ by computing softmax output for each of the $c$ classes at each time frame $m=0,1,..,M-1$. 
The number of time frames corresponding to each class $c$ is given by $T_c = \sum_{i,m} Y_{imc}$. 
During training, the base model parameters $\phi$ and $\theta$ are updated using the gradient descent algorithm as:
\begin{equation}\label{gd}
\begin{gathered}
    {[\phi,\theta]} {\leftarrow} {[\phi,\theta]} - {\alpha}{\nabla}_{\![\phi,\theta]}L_{wce}{(\,f_{[\phi,\theta]})\,}
\end{gathered}
\end{equation}
where $\alpha\in\mathbb{R}_+$ is the learning rate and $L_{wce}$ is the weighted categorical cross-entropy loss to handle the class imbalance defined as:
\begin{equation}\label{wce}
\begin{gathered}
    L_{wce} =  -\sum_{i,m,c} w_{c} Y_{imc} \log(\hat{Y}_{imc})    
\end{gathered}
\end{equation}
where $w_c \in \mathbb{R}_+$ is inversely proportional to $T_c$.
The base model $f_{[\phi,\theta]}$ is trained for $E_1$ epochs. The feature extractor layers $\phi$ will not be updated and hence remain frozen in the subsequent training steps. 


\begin{figure}
 \centering
 \includegraphics[width=7.5cm, height=5cm]{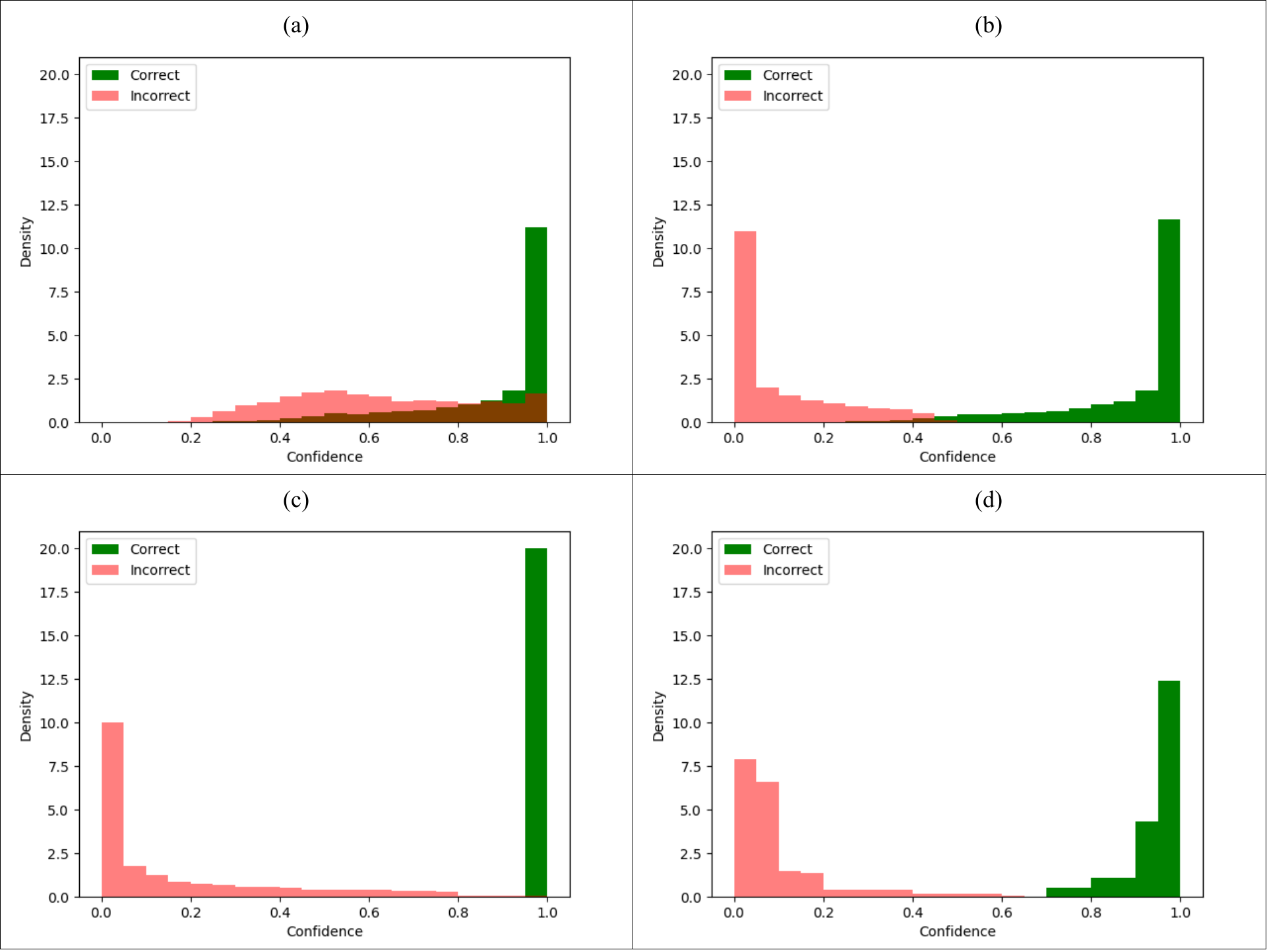}
 \caption{ Different confidence criteria derived from the output of the base model $f_{[\phi,\theta]}$. In maximum class probability (a), the correct and incorrect predictions overlap considerably. In true class probability (b), the overlap is very small and correct and incorrect predictions are well separated. Normalized true class probability (c), serves as the ground truth for training the confidence model $f_{\psi}$ where the correct predictions are assigned a value of 1 and the incorrect predictions are in the range [0,1). In (d), we show the output of the confidence model $f_{\psi}$ when trained considering (c) as the confidence criteria.}  
 \label{fig_MCP}
\end{figure}


\subsection{Confidence-model training} \label{sub:cmt}
Consider a sample $(X_i,Y_i)$. With the spectrogram $X_i$ as the input, the pre-trained base model $f_{[\phi,\theta]}$ predicts the probabilistic predictive distribution $P(\hat{Y}_i|[\phi,\theta],X_i)$ of dimension $506\times M$.
The predicted pitch class at each time frame is obtained by using argmax over the column values in that time frame. We denote these predicted pitch classes by $\hat{y}_i$ which is of dimension $M$. But in addition to the pitch classes predicted by the base model, we are also interested in the ability of the base model to recognize when its prediction is wrong, i.e., base model should be able to provide confidence values for its predictions. Ideally, for a suitable confidence criteria, a high value of confidence should indicate the correctness of the base model prediction, and a low value of confidence indicative of a wrong prediction. 

A standard confidence criteria to obtain confidence value is to consider the softmax probability corresponding to the predicted pitch class at each time frame. This is also termed as Maximum Class Probability (MCP). The MCP values for each time frame $m$ is obtained by 
\begin{equation}\label{mcp_m}
\begin{gathered}
    \underset{c}{max} \; P(\hat{Y}^{(m)}_i=c|[\phi,\theta],X_i) \\  
    = P(\hat{Y}^{(m)}_i=\hat{y}^{(m)}_i|[\phi,\theta],X_i)
\end{gathered}
\end{equation} 
It is observed that MCP leads to high confidence values for both correct and incorrect predictions, making the base model over-confident on the wrong predictions as depicted in Fig.~\ref{fig_MCP}(a). The true class at each time frame of $Y_i$ is computed similarly to that for $\hat{Y}_i$, i.e., by using argmax over the column values for each time frame in ${Y}_i$. We denote these true classes (ground truth) by $y^{*}_i$ which is of dimension $M$. For an incorrectly predicted time frame $m$, the probability associated with the true class 
$P(\hat{Y}^{(m)}_i=y^{*(m)}_i|[\phi,\theta],X_i)$ would be a low value, indicating that the base model is less confident. Therefore, True Class Probability (TCP) is a much suitable criteria than MCP for obtaining confidence as mentioned in \cite{tcp}. From Fig.~\ref{fig_MCP}(b), we observe that even for correct predictions given by the pre-trained base model $f_{[\phi,\theta]}$, the TCP value could be less than 0.50.
To overcome this shortcoming, we consider normalized TCP (TCP-n) as the confidence criteria which for every time frame $m$ is given by:  
\begin{equation}\label{tcp-n}
\begin{gathered}
        c_i^{*(m)} = \frac{P(\hat{Y}^{(m)}_i=y^{*(m)}_i|[\phi,\theta],X_i)}{P(\hat{Y}^{(m)}_i=\hat{y}^{(m)}_i|[\phi,\theta],X_i)}    
\end{gathered}
\end{equation} 
The TCP-n criteria has strong theoretical guarantee compared to TCP, as the correct predictions will be assigned a value of 1 and the incorrect predictions will be in the range [0,1) depicted in the Fig.~\ref{fig_MCP}(c).  
Since the true classes $y^{*}_{i}$ are not available when estimating confidence on target samples, therefore, the we need a model that learns the TCP-n confidence values of the training samples in $D_1^S$ dataset. 

\begin{figure}
 \centering
 \includegraphics[width=8.5cm, height=4cm]{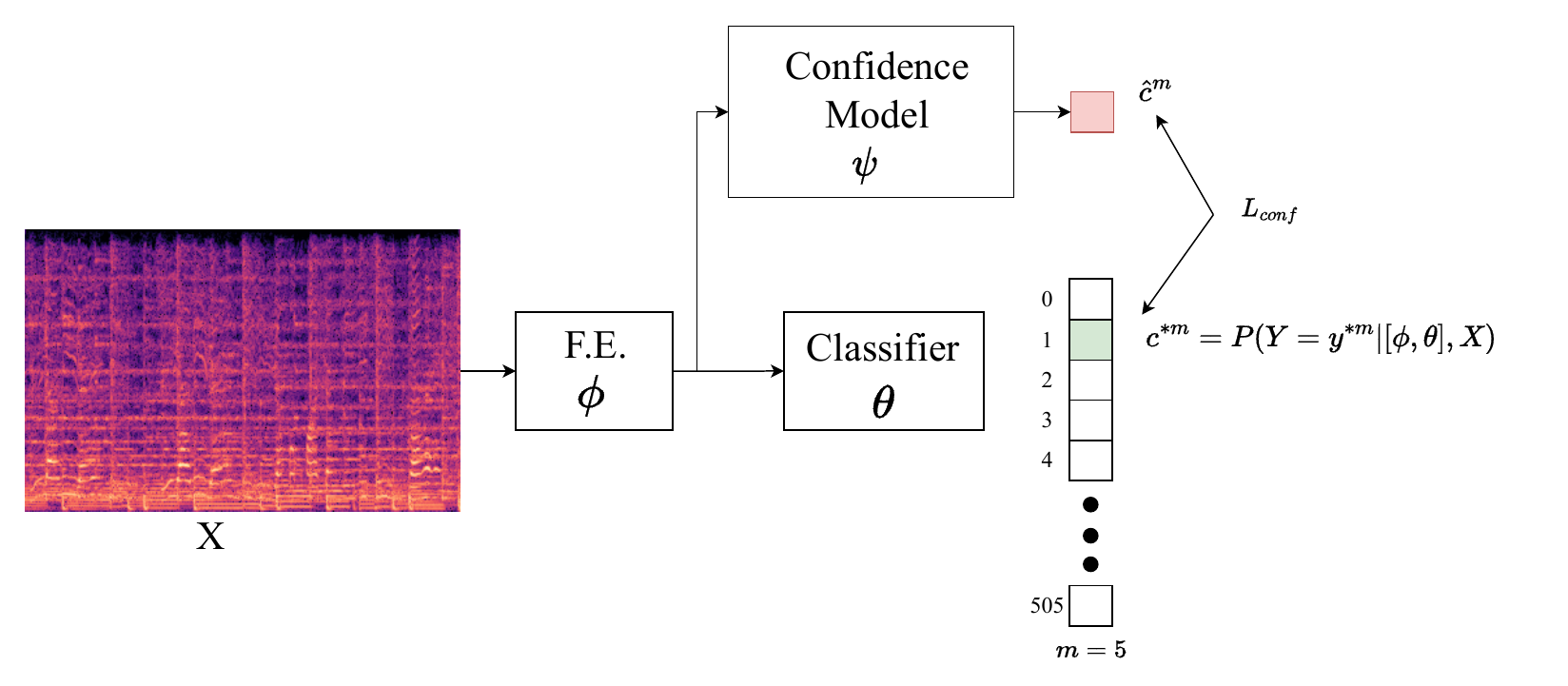}
 \caption{Here, $\phi$ and $\theta$ represents the pre-trained feature extractor (F.E) layers and classifier layer respectively. $\psi$ represents parameters of the confidence model. $L_{conf}$ is calculated at a particular time frame $m=5$. Similarly, the confidence loss is calculated at every time frame and then the confidence model is trained.}  
 \label{conf_model}
\end{figure}

To learn the TCP-n confidence values, we build a confidence model with parameters $\psi$ on top of the feature extractor layers $\phi$ of the pre-trained base model $f_{[\phi,\theta]}$. With the spectrogram $X_i$ as the input, the complex features extracted from the features extractor layers $\phi$ are fed to the confidence model with parameters $\psi$ that outputs a confidence prediction $\hat{c}_i=f_{\psi}(X_i)$. Here, $\hat{c}_i$ is of dimension $M$, i.e., a scalar confidence value $\hat{c}_i^{(m)} \in [0,1]$ is predicted at each time frame $m$. This model framework is similar to the one mentioned in \cite{tcp}.  
During training, we aim to learn the parameters $\psi$ such that $\hat{c}_i$ is close to the TCP-n confidence values $c^{*}_i$ calculated in eq.~\ref{tcp-n}. The parameters $\psi$ are updated using the gradient descent algorithm as:
\begin{equation}\label{gd_Conf}
\begin{gathered}
    \psi {\leftarrow} \psi - {\alpha}{\nabla}_{\psi}{L_{conf}{(\,f_{\psi})\,}}
\end{gathered}
\end{equation}
where $\alpha\in\mathbb{R}_+$ is the learning rate and $L_{conf}$ is the mean squared loss defined as:
\begin{equation}\label{l2_mod}
\begin{gathered}
    L_{conf} = \frac{1}{I}\sum_{i=1}^{I}(\hat{c_i}-c^{*}_i)^2
\end{gathered}
\end{equation}
The framework of the confidence model is depicted in Fig.~\ref{conf_model}. The confidence model is trained for $E_2$ epochs. Fig.~\ref{fig_MCP}(d) depicts the distribution of the output confidence values of the confidence model $f_{\psi}$ which demonstrates how well it has learned the TCP-n confidence criteria on $D_1^S$. 

\subsection{Active-Meta-Learning} \label{sub:aml}



In this section, we explain in detail how confidence-based active learning combined with model agnostic meta-learning helps to better generalize to the target domain. Given the pre-trained base model $f_{[\phi,\theta]}$ and the trained confidence model $f_\psi$, the active-meta-learning is divided into two stages as follows:

\subsubsection{\textbf{Active-Meta-training}} \label{subsubsec:amtr}
Let another source training dataset be $D_2^S = \{(X_b,Y_b)\}_{b=1}^B$, where $X_b$ and $Y_b$ are defined in the same way as $D_1^S$. Every sample in $D_2^S$ is considered an episode $b$.

For a particular episode $b$, consider two models $f_{[\phi,\theta^b]}$ and  $f_{\psi^b}$, where the layers $\phi$ are same as the feature extractor layers of the pre-trained base model, $\theta^b$ are the trainable parameters that are custom initialized by the classifier weights of the pre-trained base model, i.e., $\theta^b=\theta$ and $\psi^b$ are the trainable parameters that are custom initialized by the weights of the trained confidence model, i.e., $\psi^b=\psi$. 

With the spectrogram $X_b$ as the input, the model $f_{[\phi,\theta^b]}$ predicts an output $\hat{Y}_b = f_{[\phi,\theta^b]}(X_b)$ of the dimension $506 \times M$. We compute the predicted pitch class at every time frame $m$ of $\hat{Y}_b$ as discussed in ~\ref{sub:cmt} and denote these predicted classes by $\hat{y}_b$ which is of dimension $M$. For an episode $b$, there is a true class associated with every $m$ time frame of $Y_b$ which can be computed as in ~\ref{sub:cmt}. We denote the true classes (ground truth) by $y^*_b$ which is of dimension $M$.

The frequency of each class $c$ present in the ground truth $y^{*}_b$ is calculated by $f_c^g = \frac{n_c}{M}$, where $n_c$ is the total number of time frames corresponding to class $c$. The weight of each class is given by $w_c^g = \frac{1}{f_c^g}$, with the value being assumed as zero for the classes not present in the ground truth $y^{*}_b$. A point to note here is that we do not assign the class weights according to the class distribution across the entire source domain $D_2^S$, because these weights may not accurately reflect the current distribution of ground truth pitch classes in the episode $b$. Also upon comparing the predicted pitch classes to the ground truth pitch classes in an episode $b$, some classes maybe under or over-predicted. Therefore, we calculate the class weights dynamically for the current episode by modifying the class weights $w_c^g$ as,

\begin{figure*}[h]
 \centering
  \includegraphics[height=7cm,width=18cm]{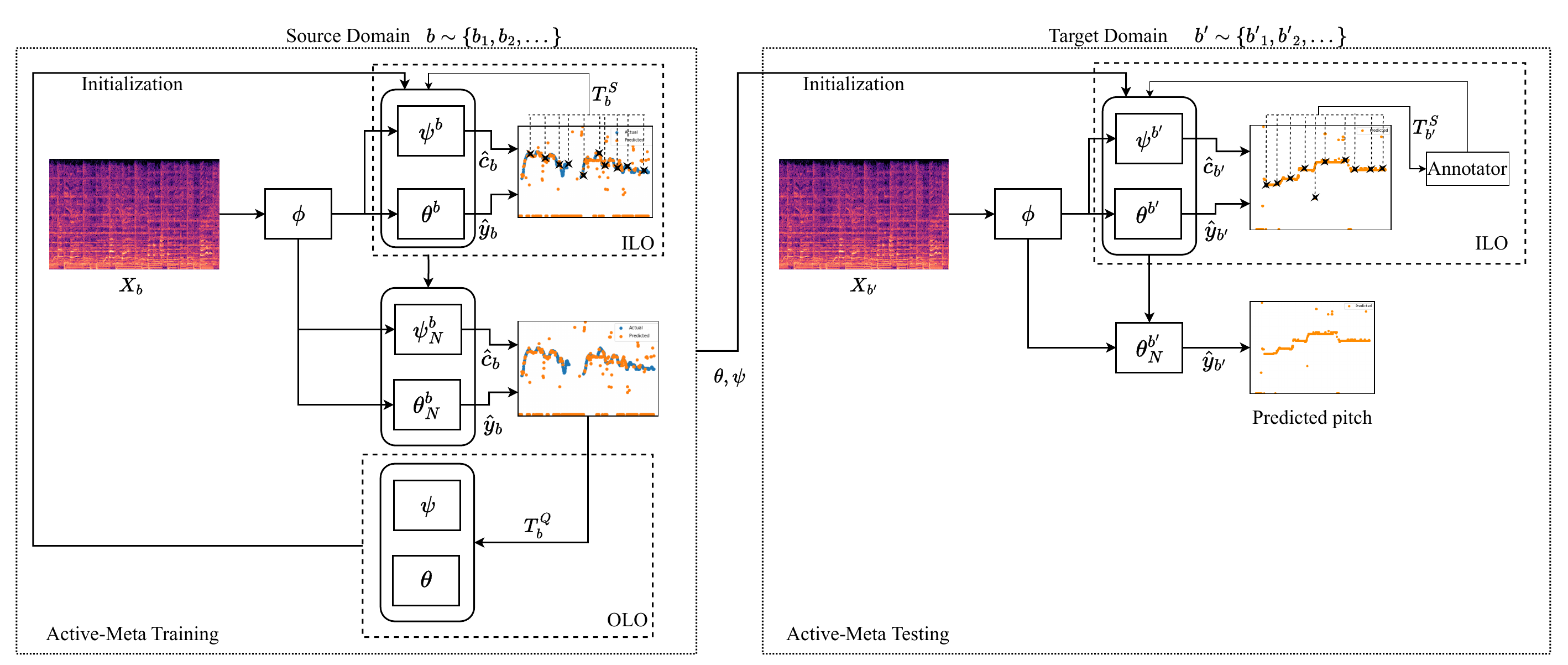}

 \caption{Active-Meta-learning framework for polyphonic melody adaptation. In active-meta-training, for an episode $b$, ILO is performed on $T_b^S$ such that the model parameters $\theta^b$ and $\psi^b$ are updated. Further OLO is performed on $T_b^Q$ to update parameters $\theta$ and $\psi$. The same procedure is repeated for all episode $b$ in source domain. In active-meta-testing, for an episode $b'$ the episode parameters are initialized as $\theta^{b'}=\theta$ and $\theta^{b'}=\psi$ and are used to adapt on $T_{b'}^S$ (single iteration of ILO) and predict on $T_{b'}^Q$.}
 \label{fig:proposed}
\end{figure*}

\begin{equation}\label{w8s}
\begin{gathered}
    {w_c^{'g}} = {w_c^g} \times e^{\lambda |\Delta w_c|} , { c=0,1,2,...,505}
\end{gathered}
\end{equation}
where $\lambda\in\mathbb{R}_+$ is a scaling factor. 
We calculate $\Delta w_c$ for all classes $c$ as $\Delta w_c = (w_c^g - w_c^p)/w_c^g$, where $w_c^p$ represents the weights of the predicted pitch classes $\hat{y}_b$ calculated in a similar way as $w_c^g$.  
Large value of $\Delta w_c$ for a particular class $c$ means that we emphasize more for the model to learn this class as it is under or over-predicted. We name this class weighting technique as meta-weighting (MW). 

With the same spectrogram $X_b$ as the input, the complex features extracted from the feature extractor layers $\phi$ are fed to the model $f_{\psi^b}$ and it predicts an output $\hat{c}_b = f_{\psi^b}(X_b)$ of dimension $M$, i.e., we have a confidence value associated with each time frame.

\begin{algorithm}[t]
  \caption{Active-Meta-training Algorithm}
  \begin{algorithmic}[1]
    \REQUIRE $\alpha$,$\beta$: learning rates
    \REQUIRE Pre-trained base model parameters [$\phi,\theta$] and confidence model parameters $\psi$ ; frozen $\phi$  
    \FOR{$E_3$ number of epochs}
        \FOR{all episodes $b$ in $D_2^S$ dataset}
          \STATE Initialize $\theta^b=\theta$ and $\psi^b=\psi$
          \STATE Compute the estimated pitch classes at each time frame from $\hat{Y}_b = f_{[\phi,\theta^b]}(X_b)$
          \STATE Compute the confidence values at each time frame as $\hat{c}_b = f_{\psi^b}(X_b)$
          \STATE Select $K$ time frames corresponding to low confidence values in $\hat{c}_b$ to form support set $T_b^S=\{m_1,...,m_K\}$ and the rest $M-K$ time frames as the query set $T_b^Q$ 
          \STATE Compute the updated parameters $\theta^b_N$ using $T_b^S$ by ILO ($N$ update steps) as in eq .~\ref{ilo}
          \STATE Calculate updated $c^{*}_b$ from $\hat{Y}_b=f_{[\phi,\theta_N^b]}(X_b)$ as in eq.~\ref{tcp-n}
          \STATE Compute the updated parameters $\psi^b_N$ using $T_b^S$ by ILO ($N$ update steps) as in eq.~\ref{ilo_conf}
          \STATE Update $\theta$ and $\psi$ using $T_b^Q$ by OLO (one update step) as in eq.~\ref{olo} and eq.~\ref{olo_conf}
        \ENDFOR
    \ENDFOR
    \STATE Obtained updated parameters $\theta$ and $\psi$ 
  \end{algorithmic}
  \label{algo1}
\end{algorithm}

\begin{algorithm}[h]
  \caption{Active-Meta-testing Algorithm ($s=1$)}
  \begin{algorithmic}[1]
    \REQUIRE $\alpha$: learning rate
    \REQUIRE $\theta$ and $\psi$ from Algorithm 1 
    \FOR{all episodes $b'$ in $D^T$ dataset}
      \STATE Initialize $\theta^{b'}=\theta$ and $\psi^{b'}=\psi$
      \STATE Compute the estimated pitch classes at each time frame from $\hat{Y}_{b'} = f_{[\phi,\theta^{b'}]}(X_{b'})$
      \STATE Compute the confidence values at each time frame as $\hat{c}_{b'} = f_{\psi^{b'}}(X_{b'})$
      \STATE From $\hat{c}_{b'}$, select $K$ least confident time frames as the support set $T_{b'}^S=\{m_1,...,m_K\}$ and provide them to the annotator to annotate and the rest $M-K$ time frame as the query set $T_{b'}^Q$ 
      \STATE Compute the updated parameters $\theta^{b'}_N$ using $T_{b'}^S$ by ILO ($N$ update steps) as in ~eq.\ref{ilo}
      \STATE Calculate updated $c^{*}_{b'}$ from $\hat{Y}_{b'}=f_{[\phi,\theta_N^{b'}]}(X_{b'})$ as in eq.~\ref{tcp-n}
      \STATE Compute the updated parameters $\psi^{b'}_N$ using $T_{b'}^S$ by ILO ($N$ update steps) as in eq.~\ref{ilo_conf} 
      \STATE From the model with updated parameters $\theta_N^{b'}$, obtain the estimated pitch classes on the query set from output $\hat{Y}_{b'}=f_{[\phi,\theta_N^{b'}]}(X_{b'})$ 
    \ENDFOR
  \end{algorithmic}
  \label{algo2}
\end{algorithm}

To create the support set $T_b^S$ for an episode $b$, we consider $K$ time frames corresponding to the low confidence values in $\hat{c}_b$. It is denoted by $T_b^S = \{m_1,m_2,...,m_K\}$. This means that the model $f_{[\phi,\theta^b]}$ is least confident at these $K$ time frames. The query set $T_b^Q$ for an episode $b$ consists of rest of the $M-K$ time frames. The classifier layer $\theta^b$ of the model $f_{[\phi,\theta^b]}$ is trained on the support set $T_b^S$ of episode $b$ and is given by the following equation:
\begin{equation}\label{ilo}
\begin{gathered}
    {\theta}_i^b = {\theta}_{i-1}^b - {\alpha}{\nabla}_{\!\theta_{i-1}^b}{L_{T_b^{S}}}({f_{[\phi,\theta_{i-1}^b]}})
\end{gathered}
\end{equation}
where ${\alpha}\in\mathbb{R}_+$ is the learning rate of the model, ${\theta}_i^b$ are the updated weights of the classifier layer of the model $f_{[\phi,\theta^b]}$ after $i$ steps. The loss ${{L}_{{T}_b^{S}}}({f_{[\phi,\theta_{i-1}^b]}})$ after ${(i-1)}$ update steps is calculated as the weighted categorical cross-entropy loss as mentioned in eq.~\ref{wce} by using the updated weights $w_c^{'g}$ calculated in eq.~\ref{w8s} for the classes corresponding to the $K$ time frames present in $T_b^S$. After $N$ update steps, the updated parameters become ${\theta}^{b}_{N}$. For the same spectrogram $X_b$, the updated model $f_{[\phi,{\theta}^{b}_{N}]}$ predicts an output $\hat{Y}_b = f_{[\phi,{\theta}^{b}_{N}]}(X_b)$ of dimension $506\times M$. We calculate the updated $c^{*}_b$ from $\hat{Y}_b$ as in eq.~\ref{tcp-n}.

Further the model $f_{\psi^b}$ is updated on the support set $T_b^S$ of episode $b$ and is given by following equation:
\begin{equation}\label{ilo_conf}
\begin{gathered}
    {\psi}_i^b = {\psi}_{i-1}^b - {\alpha}{\nabla}_{\!\psi_{i-1}^b}{L_{T_b^{S}}}({f_{\psi_{i-1}^b}})
\end{gathered}
\end{equation}
where $\alpha\in\mathbb{R_+}$ is the learning rate of the model, ${\psi}_i^b$ are the updated weights of the model $f_{\psi^b}$ after $i$ steps. The loss ${L_{T_b^{S}}}({f_{\psi_{i-1}^b}})$ after $(i-1)$ update steps is calculated as in eq.~\ref{l2_mod}, considering $\hat{c}_b$ and updated $c^{*}_b$ corresponding to the $K$ time frames in $T_b^S$. After $N$ updates, the updated parameters become $\psi_N^b$. This process of updating the models $f_{[\phi,\theta^b]}$ and $f_{\psi^b}$ on the support set is called inner-loop optimization (ILO). After ILO, the models  $f_{[\phi,\theta^b]}$ and $f_{\psi^b}$ becomes $f_{[\phi,\theta_N^b]}$ and $f_{\psi_N^b}$ respectively, where the updated parameters learn episode specific knowledge and confidence values respectively, and are used for inference on the query set ${T}_{b}^{Q}$. 

The parameters $\theta$ and $\psi$ are updated using the loss over the query set ${T}_{b}^{Q}$.
This process of updating $\theta$ and $\psi$ is called outer-loop optimization (OLO) which is expressed by,

\begin{equation}\label{olo}
\begin{gathered}
    {\theta} {\leftarrow} {\theta} - {\beta} {\nabla_{\theta}}{{L}_{{T}_{b}^{Q}}}{(\,f_{{\theta}_{N}^b})\,}
\end{gathered}
\end{equation}

\begin{equation}\label{olo_conf}
\begin{gathered}
    {\psi} {\leftarrow} {\psi} - {\beta} {\nabla_{\psi}}{{L}_{{T}_{b}^{Q}}}{(\,f_{{\psi}_{N}^b})\,}
\end{gathered}
\end{equation}

where $\beta\in\mathbb{R}_+$ is the learning rate, ${L}_{{T}_{b}^{Q}}(f_{{\theta}_{N}^b})$ is the weighted categorical cross-entropy loss calculated as mentioned in eq.~\ref{wce} by using the updated weights $w_c^{'g}$ calculated in eq.~\ref{w8s} for the classes corresponding to $M-K$ time frames in $T_b^Q$ and ${L}_{{T}_{b}^{Q}}(f_{{\psi}_{N}^b})$ is the mean-squared loss calculated as mentioned in eq.~\ref{l2_mod}, with $\hat{c}_b$ and $c^{*}_b$ corresponding to the $M-K$ time frames in $T_b^Q$.  

For an episode $b$, we perform $N$ updates on both the models in inner-loop optimization, and only one update on both the models in the outer-loop optimization. The updated model $f_{[\phi,\theta]}$ and $f_{\psi}$ for one episode is used to initialize the model $f_{\theta^b}$ and $f_{\psi^b}$ for the next episode $b$.  
The entire inner-loop and outer-loop optimization process (two-stage optimization) is repeated for all the episodes $b$ in the $D_2^S$ dataset in source domain for $E_3$ number of epochs. The final updated models $f_{[\phi,\theta]}$ and $f_{\psi}$ learn knowledge across  all episodes. The algorithm for active-meta-training is mentioned in \textbf{Algorithm \ref{algo1}}.

\subsubsection{\textbf{Active-Meta-testing}} \label{subsub:amte}
In this stage, we test the trained models $f_{[\phi,\theta]}$ and $f_{\psi}$ on the target domain. Let the target dataset be $D^T = \{X_{b'}\}_{b'=1}^{B}$, where $X_{b'}$ is the spectrogram of shape $F\times M$. 

After active-meta-training, the updated model parameters $\theta$ and $\psi$ now act as good initialization parameters for the model that adapts to the audios of different singers or genres. For a particular episode $b'$ in the target domain, the model $f_{[\phi,\theta^{b'}]}$ is initialized as $\theta^{b'}=\theta$ and the confidence model $f_{\psi^{b'}}$ is initialized as $\psi^{b'}=\psi$. With the spectrogram $X_{b'}$ as the input, the model predicts an output $\hat{Y}_{b'}=f_{[\phi,\theta^{b'}]}(X_{b'})$. We compute the predicted pitch class at every time frame of $\hat{Y}_{b'}$ as discussed in ~\ref{sub:cmt} and denote it by $\hat{y}_{b'}$. With the same spectrogram $X_{b'}$ as the input, the confidence model $f_{\psi^{b'}}$ predicts an output $\hat{c}_{b'}=f_{\psi^{b'}}({X_{b'}})$. In each iteration of inner-loop optimization, we select $K$ least confident frames from $\hat{c}_{b'}$ as the support set $T_{b'}^S$. The time frames in the support set are given to the annotator. The annotator annotates and thus provide ground truth pitch classes $y^{*}_{b'}$ to these frames. We name the process of confidence-based time frame selection for annotation as active adaptation (AA). The $N$ update steps are performed over the model parameters $\theta^{b'}$ 
on support set $T_{b'}^S$ as in eq.~\ref{ilo} where the class weights ${w'}_c^g$ for the ground truth pitch classes corresponding to the $K$ time frames in $T_{b'}^S$ are calculated as in the eq.~\ref{w8s}. After $N$ update steps, the updated parameters become ${\theta}^{b'}_{N}$. We calculate the updated $c^*_{b'}$ in the similar way as in~\ref{subsubsec:amtr}. The confidence model $f_{\psi^{b'}}$ is updated on the support set as in eq.~\ref{ilo_conf} considering $\hat{c}_{b'}$ and updated $c^{*}_{b'}$ corresponding to $K$ time frames in $T_{b'}^S$. The above ILO process is repeated for $s$ number of iterations. In this way, we adapt to $sK$ time frames. After adapting on $sK$ frames, the performance of the model with the final updated parameters $\theta^{b'}_{N}$ is finally evaluated on the query set containing $M-sK$ time frames. The results are averaged over the query set of all episodes in $D^T$ to assess the generalizability of the model. The algorithm for active-meta-testing for single iteration ($s=1$) of ILO is mentioned in \textbf{Algorithm \ref{algo2}}. The entire proposed method of active-meta-learning catering to class imbalance is denoted by w-AML. The framework for interactive melody adaptation is depicted in Fig.~\ref{fig:proposed}.

\section{HAR : Hindustani alankaar and raga dataset} \label{sec:har}
The HAR dataset created consists of 523 audio files (alankaars and ragas) of about 6.84 hours. The dataset is created by two Hindustani classical vocalists. There are 259 audio files recorded by first vocalist of about 2.6 hours. There are 264 audio files recorded by second vocalist of about 4.24 hours. 

For recording the audios, the audio setup necessitates simultaneous playback and recording. To create the dataset, the tanpura and percussion instruments are played. The simultaneous playback and recording of audio unavoidably introduce a time shift in time synchronization. Typically, the time shift in audio playback and recording remains constant for a given recording setup and equipment, so it can be mitigated through a one-time calibration and time shifting of one of the audio files to achieve a zero relative time shift.

Given the above described audio setup, audio playback is conducted via headphones so that it is not recorded in the channel recording the singing. Each vocalist determines the scale of tanpura and the BPM (beats per minute) of the percussion. The singing audio is time shifted to maintain time synchronization and the annotations are obtained by Praat~\cite{praat}. The polyphonic dataset is created by mixing the time-shifted audio with the tanpura and percussion audio to maintain time synchronicity. 

Our recording setup includes an Audio-Technica AT2020 cardioid condenser microphone and a laptop with a 12th GenIntel Core i5-12500H 12-core processor. For recording, the Audacity software~\cite{audacity} is used. 



\section{Experiments} \label{exps}
\subsection{Data}
For the melody adaptation task, we have used MIR1K{\footnote{\url{https://sites.google.com/site/unvoicedsoundseparation/mir-1k}}} as the source data $D^S$ which contains 1000 Chinese karaoke clips corresponding to 19 singers, out of which we consider 739 audios corresponding to first 14 singers in $D_1^S$ and 261 audios corresponding to remaining singers in $D_2^S$. For each dataset $D_1^S$ and $D_2^S$, we divide the train and validation data audios in the ratio 90:10, such that there are 665 train and 74 validation audios in $D_1^S$ and 235 train and 26 validation audios in $D_2^S$, i.e., a total of 900 train and 100 validation data audios in $D^S$. The total training dataset consists of about 2.2 hours of data. No data augmentation is performed. We have tested the performance of the model on the three target datasets $D_1^T$: ADC2004{\footnote{\label{note1}\url{http://labrosa.ee.columbia.edu/projects/melody/}}}, $D_2^T$: MIREX05\footref{note1} and $D_3^T$: HAR{\footnote{\url{https://zenodo.org/record/8252222}}}. The proposed model is only trained for singing voice melody, so we have selected only those test samples that contained melody sung by humans. As a result, 12 clips in ADC2004($D_1^T$) and 9 clips in MIREX05($D_2^T$) are selected. Since we divide the audios into 5-second chunks, we have a total of 1925 train and 236 validation audio chunks in $D^S$, with 1535 train and 189 validation audio chunks in $D_1^S$ and 390 train and 47 validation audio chunks in $D_2^S$. Further, we have 43, 60 and 2847 audio chunks in $D_1^T$, $D_2^T$ and $D_3^T$ respectively. 

\subsection{Experiment setting}
In this paper, we employ a basic deep CNN model as the base model to carry out the melody extraction task. The base model consists of 4 convolutional layers having [64,128,192,256] filters each using kernels of size $5 \times 5$ with batch normalization and ReLU activation followed by a dense layer having 512 nodes with ReLU activation and a Timedistributed classifier layer with 506 nodes with softmax activation. The calculated spectrogram is of dimension $F \times M$, where $F=513$ frequency bins and $M=500$ time frames. The confidence model 
consists of a dense layer having 256 nodes with ReLU activation followed by a layer of single node with sigmoid activation.

In the proposed experiment, we pre-train the base model on train data in $D_1^S$ for $E_1=450$ epochs as in eq.~\ref{gd} with a learning rate of $1\times10^{-5}$. After pre-training the base model, we train the confidence model on same $D_1^S$ dataset for $E_2=200$ epochs as in eq.~\ref{gd_Conf} with a learning rate of $1\times10^{-5}$. We perform active-meta-training on train data in $D_2^S$. For ILO, we consider $K=10$ time frames as the support set. We update the weights of the classifier and confidence model with $N=10$ inner-loop updates on the support set as in eq.~\ref{ilo} and eq.~\ref{ilo_conf} with the learning rate of $1\times10^{-5}$, respectively. The same learning rate is used for OLO. We consider $\lambda=0.2$ in eq.~\ref{w8s}. The entire two-stage optimization process is repeated for $E_3=400$ epochs. During active-meta-testing on each of the target domain datasets $D_1^T$,$D_2^T$and $D_3^T$, we consider a single iteration ($s=1$) of ILO by selecting $K=10$ time frames as support set. We update the weights of the classifier and confidence model with $N=10$ inner-loop updates on the support set as in eq.~\ref{ilo} and eq.~\ref{ilo_conf} with a learning rate of $1\times 10^{-5}$ respectively and predict on the rest $M-K=490$ time frames.

We compare the performance of our proposed algorithm with the baseline algorithms. To maintain the valid comparison, we keep the same source and target domains across all the baseline experiments.
We categorize the baseline experiments into two categories: Non-adaptive and Adaptive experiments. We explain the experiments as follows:
\begin{enumerate}
    \item \textbf{Non-adaptive experiment:} 
    We perform classical training (CT) on the base model. We pre-train the base model on the train data in $D^S$ for 450 epochs by using eq.~\ref{gd} with a learning rate of $1\times10^{-5}$. No adaptation on the target data is performed. The trained model is used to evaluate the performance on the validation dataset and target datasets - $D_1^T$,$D_2^T$ and $D_3^T$. The other non-adaptive baselines include Patch-based CNN~\cite{patch-basedcnn}, Attention Network~\cite{attention} and SegNet~\cite{en-decoder}. We have obtained the results of these experiments on the audios in validation data in $D^S$ and target datasets $D_1^T$,$D_2^T$ and $D_3^T$ by downloading their online source codes and compiling the results on our dataset configuration. The baselines are also trained using CT.
    
    \item \textbf{Adaptive experiments:} We compare our proposed algorithm with the following adaptive baselines:
    \begin{itemize}
        \item Fine-Tuning (FT)~\cite{ft}: We pre-train the base model in the same way as in CT. We do not consider an additional confidence model in this method. After pre-training, we update the classifier by adapting on randomly selected $K=10$ time frames of audio episodes in the each target dataset. We name this as random adaptation (RA). We evaluate the adapted model on the rest of the $M-K=490$ time frames for each audios in target datasets. 
        \item MAML~\cite{maml-finn}: We pre-train the base model on the train data in $D_1^S$ for 450 epochs as in eq.~\ref{gd} with a learning rate of $1\times10^{-5}$. We do not consider an additional confidence model. During meta-training on the train data in $D_2^S$, for ILO we consider random $K=10$ time frames as the support set and update the weights of the classifier with $N=10$ inner-loop updates as in eq.~\ref{ilo} with a learning rate of $1\times10^{-5}$. The same learning rate is used for OLO. The entire two-stage optimization process is repeated for 450 epochs. Note that during meta-training no class weights are considered as in original MAML~\cite{maml-finn}, i.e., no MW. During meta-testing on the target datasets $D_1^T$,$D_2^T$ and $D_3^T$, we perform RA by considering $K=10$ time frames as the support set. We evaluate the adapted model on the query set, i.e., rest of $M-K=490$ frames for each episode in the target datasets.       
    \end{itemize}        
\end{enumerate}
The performance metrics considered are raw pitch accuracy (RPA), raw chroma accuracy (RCA) and overall accuracy (OA). All these metrics are computed by using a standard mir-eval\cite{mir_eval} library with a pitch detection tolerance of 50 cents. We further perform the proposed w-AML with different values of support set size, i.e., $K=10,15,20$ to understand the effect of increasing support set size.

  

\begin{table*}[!t]
\caption{Performance metrics with the base model used by us and other baseline methods on the validation dataset (source dataset) and the three target datasets. All models are trained using CT. No adaptation used. \label{tab:non-adaptive}}
\centering
  \begin{tabular}{|*{13}{c|}} \cline{1-13} 
  \hline
  \multicolumn{1}{|c}{\multirow{2}{*}{\textbf{Experiments}}} & \multicolumn{3}{|c|}{\textbf{MIR1K-val}} & \multicolumn{3}{|c|}{\textbf{ADC2004}} & \multicolumn{3}{c|}{\textbf{MIREX05}} & \multicolumn{3}{c|}{\textbf{HAR}} \\ \cline{2-13}
  \multicolumn{1}{|c|}{} & \textbf{RPA} & \textbf{RCA}  & \textbf{OA} & \textbf{RPA} & \textbf{RCA}  & \textbf{OA} & \textbf{RPA} & \textbf{RCA} & \textbf{OA}  & \textbf{RPA} & \textbf{RCA} & \textbf{OA} \\
  \hline  
  Patch-based CNN\cite{patch-basedcnn} & 86.12 & 86.25 & 86.88 & 76.30 & 76.70  & 78.40 & 74.30 & 81.20  & 82.20 & 62.20  & 60.60 & 61.89\\
  \hline
  Attention Network\cite{attention} & 88.67 & 88.34 & 89.30 &76.30 & 76.50 & 77.40 & 77.80 & 77.80 & 84.40  & 65.40 & 66.32 & 66.55\\
  \hline
  SegNet\cite{en-decoder} & \textbf{89.10} & \textbf{89.16} & \textbf{90.10} &\textbf{82.70} & \textbf{84.90} & \textbf{81.60} & 78.40 & 79.70 & 78.60 & 68.34  & 69.32 & 65.63\\
  \hline
  \textbf{Our base model}& 88.64 & 88.90 & 88.45 & 79.26 & 80.55 & 79.90 & \textbf{81.88} & \textbf{82.15} & \textbf{81.30} & \textbf{75.43} & \textbf{76.70} & \textbf{75.90}\\
  \hline
  \end{tabular}
\end{table*}

\begin{table*}[!t]
\caption{Performance metrics with adaptive methods on the three target datasets. Here, MW, AA and RA stand for meta-weighting, active adaptation and random adaptation, respectively. \label{tab:adaptive}}
\centering
  \begin{tabular}{|*{13}{c|}} \cline{1-13} 
  \hline
  \multicolumn{4}{|c|}{\textbf{Experiments}} & \multicolumn{3}{|c|}{\textbf{ADC2004}} & \multicolumn{3}{c|}{\textbf{MIREX05}} & \multicolumn{3}{c|}{\textbf{HAR}} \\ \cline{1-13}
  \textbf{Method} & \textbf{MW} & \textbf{AA} & \textbf{RA} & \textbf{RPA} & \textbf{RCA}  & \textbf{OA} & \textbf{RPA} & \textbf{RCA} & \textbf{OA}  & \textbf{RPA} & \textbf{RCA} & \textbf{OA} \\
  \hline  
   FT~\cite{ft} & - & \xmark & \cmark &80.34 & 81.45 & 80.98 & 81.16 & 81.98 & 82.10 & 76.45 & 77.10 & 76.88\\
  \hline
  MAML~\cite{maml-finn} & \xmark & \xmark & \cmark &81.10 & 82.56 & 81.41 & 83.16 & 84.57 & 83.28 & 77.70 & 78.12 & 78.10\\     
  \hline
  \textbf{w-AML(Ours)} & \cmark & \cmark & \xmark & \textbf{86.40} & \textbf{87.01} & \textbf{86.15} & \textbf{87.23} & \textbf{88.15} & \textbf{87.80} & \textbf{80.60} & \textbf{80.99} & \textbf{81.45}\\
  \hline
  \end{tabular}
\end{table*}

\begin{table*}[!t]
\caption{Raw pitch accuracy on three target datasets for different support set size $K$ \label{tab:diffkrpa}}
\centering
  \begin{tabular}{|*{10}{c|}} \cline{1-10} 
  \hline
  \multicolumn{1}{|c}{\multirow{2}{*}{\textbf{Experiment}}} & \multicolumn{3}{|c|}{\textbf{ADC2004}} & \multicolumn{3}{c|}{\textbf{MIREX05}} & \multicolumn{3}{c|}{\textbf{HAR}} \\ \cline{2-10}
  \multicolumn{1}{|c|}{} & \textbf{$K=10$} & \textbf{$K=15$}  & \textbf{$K=20$} & \textbf{$K=10$} & \textbf{$K=15$}  & \textbf{$K=20$}  & \textbf{$K=10$} & \textbf{$K=15$}  & \textbf{$K=20$} \\
  \hline  
  \textbf{w-AML(Ours)} & 86.40 & 87.20 & \textbf{88.95} & 87.23 & 88.45 & \textbf{89.99} & 80.60 & 81.55 & \textbf{82.01}\\
  \hline
  \end{tabular}
\end{table*}

\begin{table*}[!t]
\caption{Ablation study on the three target datasets. Here, MW, AA and RA stand for meta-weighting, active adaptation and random adaptation, respectively. \label{tab:ablation}}
\centering
  \begin{tabular}{|*{13}{c|}} \cline{1-13} 
  \hline
  \multicolumn{4}{|c|}{\textbf{Experiments}} & \multicolumn{3}{|c|}{\textbf{ADC2004}} & \multicolumn{3}{c|}{\textbf{MIREX05}} & \multicolumn{3}{|c|}{\textbf{HAR}} \\ \cline{1-13}
  \textbf{Method} & \textbf{MW} & \textbf{AA} & \textbf{RA} & \textbf{RPA} & \textbf{RCA}  & \textbf{OA} & \textbf{RPA} & \textbf{RCA} & \textbf{OA}  & \textbf{RPA} & \textbf{RCA} & \textbf{OA} \\
  \hline  
   w-MAML & \cmark & \xmark & \cmark & 83.50 & 84.81 & 84.99 & 85.39 & 86.88 & 85.95 & 77.34  & 78.45 & 77.14 \\     
  \hline
  AML & \xmark & \cmark & \xmark & 81.32 & 82.56 & 81.99 & 82.12 & 83.88 & 81.80 & 75.78 & 76.55 & 75.98\\   
  \hline
\textbf{w-AML(Ours)} & \cmark & \cmark & \xmark & \textbf{86.40} & \textbf{87.01} & \textbf{86.15} & \textbf{87.23} & \textbf{88.15} & \textbf{87.80} & \textbf{80.60} & \textbf{80.99} & \textbf{81.45}\\
  \hline
  \end{tabular}
\end{table*}


\begin{figure}[!t]
 \centering
  \includegraphics[height=5.5cm,width=7.5cm]{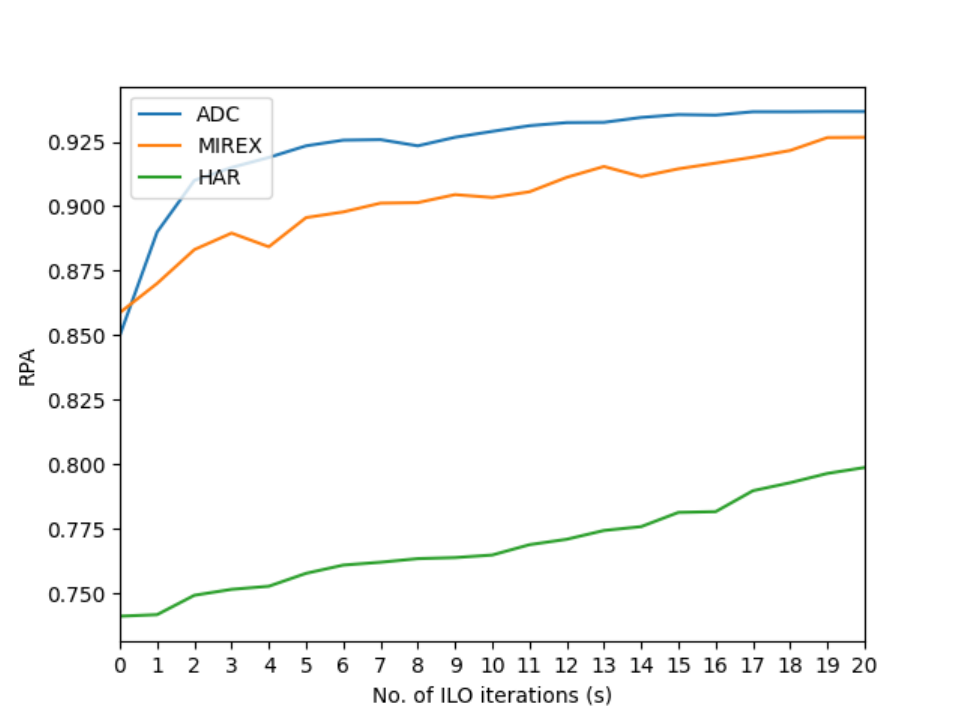}    
 \caption{RPA on the query set of size $(M-sK)$ vs $s$ for a typical episode from the three target datasets. Here, $s=0$ means no adaptation is performed.}
 \label{compadapt}
\end{figure}

\section{Results and discussions}
\label{result}
Table~\ref{tab:non-adaptive} depicts the comparison of the performance of each method on the validation data from source domain and the different target domain datasets. We observe that the performance of each method trained using CT degrades in extracting melody from the target domains. This illustrates how model performance is adversely affected by domain shift. HAR dataset consists of hindustani classical music samples which differs from those of MIR1K dataset in terms of singing styles, musicological structure and background instruments. The effect of this large domain shift clearly reflects in poor performance of all models on HAR dataset. Further, we observe that our base model performs better on the target domains MIREX05 and HAR. Even though our base model does not perform well on ADC2004 data, we see that with various domain adaptation methods the performance of the base model can further be improved as shown in Table~\ref{tab:adaptive}. 

Table~\ref{tab:adaptive} depicts the different domain adaptation methods applied to our base model and we observe that each method improves the results on the target domain data as compared to Table~\ref{tab:non-adaptive}. Further, the table shows the comparison of our proposed w-AML method to the other adaptive baseline methods, i.e., FT and MAML. Even though both the baselines use RA, MAML performs better than FT. This observation is in line with the findings in previous literature~\cite{maml-finn}. Also, the original MAML algorithm does not cater to class imbalance problem. The proposed w-AML outperforms both the baselines as it addresses class imbalance and uses AA instead of RA.

From Table~\ref{tab:diffkrpa} we observe that the performance of the model increases as we increase the support set size because the model is able to adapt on more number of time frames. This is justified because the model performance increases when large amount of annotated data is given to the model. But our aim is to obtain a robust model by adapting on as minimal a number of least confident frames as possible, hence we consider $K=10$. The adaptation performance of our w-AML method is shown in Fig.~\ref{compadapt}. The plot shows gradual improvement in RPA on the query set of a typical episode from three target datasets after $s$ iterations of inner-loop optimization on support sets.

\section{Ablation studies}
\label{ablation}
To study the effect of various components of the proposed method, we perform ablation studies. The The following ablation experiments are carried out:
\begin{itemize}
    \item w-MAML: In this experiment, we modify the original MAML algorithm~\cite{maml-finn} such that during meta-training, it caters to class imbalance and uses MW to update classifier layer. No separate confidence model is trained, hence RA is performed. The rest of the method is same as described in MAML under adaptive experiments. 
    \item AML: In this experiment, we use our proposed w-AML method with a slight modification that during active-meta-training and active-meta-testing, class imbalance is not catered, hence no MW is applied.     
\end{itemize}
The results of these ablation experiments are presented in Table~\ref{tab:ablation}. In w-MAML, although the class imbalance is catered, the performance degrades because RA is performed as compared to AA in w-AML. In AML, although AA is performed, the performance degrades because it does not cater to class imbalance as compared to w-AML.


\section{Conclusion}
\label{ending}
In this work, we have studied the problem of domain shift in polyphonic melody extraction. To handle this problem, we have proposed a novel interactive melody adaptation algorithm based on active-meta-learning, that also handles severe class imbalance problem in audio data. The proposed algorithm is model-agnostic and can be applied to any non-adaptive model to further improve the performance. The algorithm can be used for speeding up the annotation of audios in different domains that can be used in various downstream applications of melody extraction.  



\bibliographystyle{IEEEtran}
\bibliography{bibliography}

\begin{thebibliography}{10}
\providecommand{\url}[1]{#1}
\csname url@samestyle\endcsname
\providecommand{\newblock}{\relax}
\providecommand{\bibinfo}[2]{#2}
\providecommand{\BIBentrySTDinterwordspacing}{\spaceskip=0pt\relax}
\providecommand{\BIBentryALTinterwordstretchfactor}{4}
\providecommand{\BIBentryALTinterwordspacing}{\spaceskip=\fontdimen2\font plus
\BIBentryALTinterwordstretchfactor\fontdimen3\font minus
  \fontdimen4\font\relax}
\providecommand{\BIBforeignlanguage}[2]{{%
\expandafter\ifx\csname l@#1\endcsname\relax
\typeout{** WARNING: IEEEtran.bst: No hyphenation pattern has been}%
\typeout{** loaded for the language `#1'. Using the pattern for}%
\typeout{** the default language instead.}%
\else
\language=\csname l@#1\endcsname
\fi
#2}}
\providecommand{\BIBdecl}{\relax}
\BIBdecl

\bibitem{musicrecom}
K.~Chen, B.~Liang, X.~Ma, and M.~Gu, ``Learning audio embeddings with user
  listening data for content-based music recommendation,'' in \emph{ICASSP
  2021-2021 IEEE International Conference on Acoustics, Speech and Signal
  Processing (ICASSP)}.\hskip 1em plus 0.5em minus 0.4em\relax IEEE, 2021, pp.
  3015--3019.

\bibitem{coversong}
X.~Du, K.~Chen, Z.~Wang, B.~Zhu, and Z.~Ma, ``Bytecover2: Towards
  dimensionality reduction of latent embedding for efficient cover song
  identification,'' in \emph{ICASSP 2022-2022 IEEE International Conference on
  Acoustics, Speech and Signal Processing (ICASSP)}.\hskip 1em plus 0.5em minus
  0.4em\relax IEEE, 2022, pp. 616--620.

\bibitem{musicgen}
K.~Chen, C.-i. Wang, T.~Berg-Kirkpatrick, and S.~Dubnov, ``Music sketchnet:
  Controllable music generation via factorized representations of pitch and
  rhythm,'' \emph{arXiv preprint arXiv:2008.01291}, 2020.

\bibitem{voicesep}
Y.~Ikemiya, K.~Yoshii, and K.~Itoyama, ``Singing voice analysis and editing
  based on mutually dependent f0 estimation and source separation,'' in
  \emph{2015 IEEE International Conference on Acoustics, Speech and Signal
  Processing (ICASSP)}.\hskip 1em plus 0.5em minus 0.4em\relax IEEE, 2015, pp.
  574--578.

\bibitem{patch-basedcnn}
L.~Su, ``Vocal melody extraction using patch-based cnn,'' in \emph{2018 IEEE
  International Conference on Acoustics, Speech and Signal Processing
  (ICASSP)}.\hskip 1em plus 0.5em minus 0.4em\relax IEEE, 2018, pp. 371--375.

\bibitem{da1}
S.~Ben-David, J.~Blitzer, K.~Crammer, and F.~Pereira, ``Analysis of
  representations for domain adaptation,'' \emph{Advances in neural information
  processing systems}, vol.~19, 2006.

\bibitem{alsurvey}
P.~Ren, Y.~Xiao, X.~Chang, P.-Y. Huang, Z.~Li, B.~B. Gupta, X.~Chen, and
  X.~Wang, ``A survey of deep active learning,'' \emph{ACM computing surveys
  (CSUR)}, vol.~54, no.~9, pp. 1--40, 2021.

\bibitem{maml-finn}
C.~Finn, P.~Abbeel, and S.~Levine, ``Model-agnostic meta-learning for fast
  adaptation of deep networks,'' in \emph{International conference on machine
  learning}.\hskip 1em plus 0.5em minus 0.4em\relax PMLR, 2017, pp. 1126--1135.

\bibitem{fsl1}
O.~Vinyals, C.~Blundell, T.~Lillicrap, D.~Wierstra \emph{et~al.}, ``Matching
  networks for one shot learning,'' \emph{Advances in neural information
  processing systems}, vol.~29, 2016.

\bibitem{sp1}
V.~Rao and P.~Rao, ``Vocal melody extraction in the presence of pitched
  accompaniment in polyphonic music,'' \emph{IEEE Transactions on Audio,
  Speech, and Language Processing}, vol.~18, no.~8, pp. 2145--2154, 2010.

\bibitem{aroraonline}
V.~Arora and L.~Behera, ``On-line melody extraction from polyphonic audio using
  harmonic cluster tracking,'' \emph{IEEE transactions on audio, speech, and
  language processing}, vol.~21, no.~3, pp. 520--530, 2012.

\bibitem{salamonphd}
J.~J. Salamon \emph{et~al.}, ``Melody extraction from polyphonic music
  signals,'' Ph.D. dissertation, Universitat Pompeu Fabra, 2013.

\bibitem{aud-sym-tl}
W.~T. Lu, L.~Su \emph{et~al.}, ``Vocal melody extraction with semantic
  segmentation and audio-symbolic domain transfer learning.'' in \emph{ISMIR},
  2018, pp. 521--528.

\bibitem{deepsalience}
R.~M. Bittner, B.~McFee, J.~Salamon, P.~Li, and J.~P. Bello, ``Deep salience
  representations for f0 estimation in polyphonic music.'' in \emph{ISMIR},
  2017, pp. 63--70.

\bibitem{en-decoder}
T.-H. Hsieh, L.~Su, and Y.-H. Yang, ``A streamlined encoder/decoder
  architecture for melody extraction,'' in \emph{ICASSP 2019-2019 IEEE
  International Conference on Acoustics, Speech and Signal Processing
  (ICASSP)}.\hskip 1em plus 0.5em minus 0.4em\relax IEEE, 2019, pp. 156--160.

\bibitem{jdc}
S.~Kum and J.~Nam, ``Joint detection and classification of singing voice melody
  using convolutional recurrent neural networks,'' \emph{Applied Sciences},
  vol.~9, no.~7, p. 1324, 2019.

\bibitem{attention}
S.~Yu, X.~Sun, Y.~Yu, and W.~Li, ``Frequency-temporal attention network for
  singing melody extraction,'' in \emph{ICASSP 2021-2021 IEEE International
  Conference on Acoustics, Speech and Signal Processing (ICASSP)}.\hskip 1em
  plus 0.5em minus 0.4em\relax IEEE, 2021, pp. 251--255.

\bibitem{uda1}
B.~Sun, J.~Feng, and K.~Saenko, ``Return of frustratingly easy domain
  adaptation,'' in \emph{Proceedings of the AAAI conference on artificial
  intelligence}, vol.~30, no.~1, 2016.

\bibitem{ssda1}
J.~Li, G.~Li, Y.~Shi, and Y.~Yu, ``Cross-domain adaptive clustering for
  semi-supervised domain adaptation,'' in \emph{Proceedings of the IEEE/CVF
  Conference on Computer Vision and Pattern Recognition}, 2021, pp. 2505--2514.

\bibitem{sdt}
E.~Tzeng, J.~Hoffman, T.~Darrell, and K.~Saenko, ``Simultaneous deep transfer
  across domains and tasks,'' in \emph{Proceedings of the IEEE international
  conference on computer vision}, 2015, pp. 4068--4076.

\bibitem{unifiedsda}
S.~Motiian, M.~Piccirilli, D.~A. Adjeroh, and G.~Doretto, ``Unified deep
  supervised domain adaptation and generalization,'' in \emph{Proceedings of
  the IEEE international conference on computer vision}, 2017, pp. 5715--5725.

\bibitem{neuralembed}
Z.~Wang, B.~Du, and Y.~Guo, ``Domain adaptation with neural embedding
  matching,'' \emph{IEEE transactions on neural networks and learning systems},
  vol.~31, no.~7, pp. 2387--2397, 2019.

\bibitem{dsne}
X.~Xu, X.~Zhou, R.~Venkatesan, G.~Swaminathan, and O.~Majumder, ``d-sne: Domain
  adaptation using stochastic neighborhood embedding,'' in \emph{Proceedings of
  the IEEE/CVF Conference on Computer Vision and Pattern Recognition}, 2019,
  pp. 2497--2506.

\bibitem{imagenet}
A.~Krizhevsky, I.~Sutskever, and G.~E. Hinton, ``Imagenet classification with
  deep convolutional neural networks,'' \emph{Communications of the ACM},
  vol.~60, no.~6, pp. 84--90, 2017.

\bibitem{fsl2}
S.~Ravi and H.~Larochelle, ``Optimization as a model for few-shot learning,''
  2016.

\bibitem{modelini}
C.~Finn, P.~Abbeel, and S.~Levine, ``Model-agnostic meta-learning for fast
  adaptation of deep networks,'' in \emph{International conference on machine
  learning}.\hskip 1em plus 0.5em minus 0.4em\relax PMLR, 2017, pp. 1126--1135.

\bibitem{uncen1}
D.~D. Lewis, ``A sequential algorithm for training text classifiers:
  Corrigendum and additional data,'' in \emph{Acm Sigir Forum}, vol.~29,
  no.~2.\hskip 1em plus 0.5em minus 0.4em\relax ACM New York, NY, USA, 1995,
  pp. 13--19.

\bibitem{uncen2}
X.~Li and Y.~Guo, ``Adaptive active learning for image classification,'' in
  \emph{Proceedings of the IEEE conference on computer vision and pattern
  recognition}, 2013, pp. 859--866.

\bibitem{diversity1}
O.~Sener and S.~Savarese, ``Active learning for convolutional neural networks:
  A core-set approach,'' \emph{arXiv preprint arXiv:1708.00489}, 2017.

\bibitem{diversity2}
H.~T. Nguyen and A.~Smeulders, ``Active learning using pre-clustering,'' in
  \emph{Proceedings of the twenty-first international conference on Machine
  learning}, 2004, p.~79.

\bibitem{expected1}
N.~Roy and A.~McCallum, ``Toward optimal active learning through monte carlo
  estimation of error reduction,'' \emph{ICML, Williamstown}, vol.~2, pp.
  441--448, 2001.

\bibitem{expected2}
B.~Settles, M.~Craven, and S.~Ray, ``Multiple-instance active learning,''
  \emph{Advances in neural information processing systems}, vol.~20, 2007.

\bibitem{unal1}
D.~D. Lewis, ``A sequential algorithm for training text classifiers:
  Corrigendum and additional data,'' in \emph{Acm Sigir Forum}, vol.~29,
  no.~2.\hskip 1em plus 0.5em minus 0.4em\relax ACM New York, NY, USA, 1995,
  pp. 13--19.

\bibitem{ecp}
K.~Wang, D.~Zhang, Y.~Li, R.~Zhang, and L.~Lin, ``Cost-effective active
  learning for deep image classification,'' \emph{IEEE Transactions on Circuits
  and Systems for Video Technology}, vol.~27, no.~12, pp. 2591--2600, 2016.

\bibitem{gal}
Y.~Gal, R.~Islam, and Z.~Ghahramani, ``Deep bayesian active learning with image
  data,'' in \emph{International conference on machine learning}.\hskip 1em
  plus 0.5em minus 0.4em\relax PMLR, 2017, pp. 1183--1192.

\bibitem{mcmc}
Y.~Gal and Z.~Ghahramani, ``Dropout as a bayesian approximation: Representing
  model uncertainty in deep learning,'' in \emph{international conference on
  machine learning}.\hskip 1em plus 0.5em minus 0.4em\relax PMLR, 2016, pp.
  1050--1059.

\bibitem{tcp}
C.~Corbi{\`e}re, N.~Thome, A.~Bar-Hen, M.~Cord, and P.~P{\'e}rez, ``Addressing
  failure prediction by learning model confidence,'' \emph{Advances in Neural
  Information Processing Systems}, vol.~32, 2019.

\bibitem{praat}
P.~Boersma, ``Praat, a system for doing phonetics by computer,'' \emph{Glot.
  Int.}, vol.~5, no.~9, pp. 341--345, 2001.

\bibitem{audacity}
B.~Li, J.~A. Burgoyne, and I.~Fujinaga, ``Extending audacity for audio
  annotation.'' in \emph{ISMIR}, 2006, pp. 379--380.

\bibitem{ft}
S.~K. Jha, M.~Kumar, V.~Arora, S.~N. Tripathi, V.~M. Motghare, A.~Shingare,
  K.~A. Rajput, and S.~Kamble, ``Domain adaptation-based deep calibration of
  low-cost pm\textsubscript{2.5} sensors,'' \emph{IEEE Sensors Journal},
  vol.~21, no.~22, pp. 25\,941--25\,949, 2021.

\bibitem{mir_eval}
C.~Raffel, B.~McFee, E.~J. Humphrey, J.~Salamon, O.~Nieto, D.~Liang, D.~P.
  Ellis, and C.~C. Raffel, ``mir\_eval: A transparent implementation of common
  mir metrics,'' in \emph{In Proceedings of the 15th International Society for
  Music Information Retrieval Conference, ISMIR}.\hskip 1em plus 0.5em minus
  0.4em\relax Citeseer, 2014.

\end{thebibliography}

\end{document}